\documentclass[aps,pre,twocolumn,showpacs,floatfix,superscriptaddress]{revtex4-1}
\usepackage{graphicx,amsfonts,amssymb,amsmath,hyperref}
\usepackage{textcomp}
\usepackage{esint}

\newif\ifhyper
\hypertrue
\ifhyper       
\hypersetup{        
  citecolor = {green},
  colorlinks = {true}, 
  urlcolor = {blue} 
}

\begin{document}

\def\Tkt{T_{\rm KT}}
\def\aopt{\alpha_{\rm opt}}
\def\trhor{\trho_{\rm r}}


\graphicspath{{./figures_submit/}}
\def\rhoeq{\hat\rho_{\rm eq}}

\newcommand{\marge}[1]{\marginpar{\scriptsize #1}}
\newcommand{\remarque}[1]{\marginpar{\scriptsize Remarque}{\it [#1]}}
\newcommand{\new}[1]{{\bf #1}}
\newcommand{\red}[1]{\textcolor{red}{#1}}
\newcommand{\blue}[1]{\textcolor{blue}{#1}}
\newlength{\textlarg}
\newcommand{\barre}[1]{%
   \settowidth{\textlarg}{#1}
   #1\hspace{-\textlarg}\rule[0.5ex]{\textlarg}{0.5pt}}
\newcommand{\barred}[1]{%
   \settowidth{\textlarg}{#1}
   \red{#1}\hspace{-\textlarg}\rule[0.5ex]{\textlarg}{0.5pt}}
\newcommand{\barblue}[1]{%
   \settowidth{\textlarg}{#1}
   \blue{#1}\hspace{-\textlarg}\rule[0.5ex]{\textlarg}{0.5pt}}

\def\beq{\begin{equation}}
\def\eeq{\end{equation}}
\def\bleq{\begin{eqnarray}}
\def\eleq{\end{eqnarray}} 
\def\bfig{\begin{figure}}
\def\efig{\end{figure}}
\def\bline{\begin{multline}}
\def\eline{\end{multline}}
\def\bremark{\begin{quotation} \noindent \small }
\def\eremark{\end{quotation}}
\def\llbrace{\left\lbrace}
\def\rrbrace{\right\rbrace}
\def\lbraket{\left[}
\def\rbraket{\right]}
\def\llangle{\left\langle}
\def\rrangle{\right\rangle} 

\newcommand{\Tr}{{\rm Tr}} 
\newcommand{\tr}{{\rm tr}} 
\newcommand{\sgn}{{\rm sgn}} 
\newcommand{\mean}[1]{\langle #1 \rangle}
\newcommand{\commu}[2]{[#1,#2]} 
\newcommand{\bra}[1]{\langle#1|}
\newcommand{\ket}[1]{|#1\rangle}
\newcommand{\braket}[2]{\langle #1|#2\rangle}
\newcommand{\dbraket}[3]{\langle #1|#2|#3\rangle}
\newcommand{\tens}[1]{\overleftrightarrow{#1}}  
\newcommand{\vac}{|{\rm vac}\rangle} 
\def\bravac{\langle{\rm vac}|}
\newcommand{\const}{{\rm const}} 
\newcommand{\atanh}{\,{\rm atanh}}

\newcommand{\ie}{i.e. }
\newcommand{\iet}{i.e.}
\newcommand{\eg}{e.g. }
\newcommand{\cc}{{\rm c.c.}} 
\newcommand{\hc}{{\rm h.c.}} 
\def\etal{{\it et al. }}

\newcommand{\jhatbf}{\hat {\textbf \j}} 
\newcommand{\Jhatbf}{\hat {\textbf \J}} 
\newcommand{\jhat}{\hat {\jmath}} 
\newcommand{\Jhat}{\hat {J}} 
\newcommand{\jbf}{\textbf j}
\newcommand{\Jbf}{\textbf J}

\def\chibf{\boldsymbol{\chi}}
\def\down{\downarrow}
\def\eps{\epsilon}
\def\gam{\gamma} 
\def\phibf{\boldsymbol{\phi}}
\def\varphibf{\boldsymbol{\varphi}}
\def\varphibfs{\boldsymbol{\varphi}_<}
\def\varphibfl{\boldsymbol{\varphi}_>}
\def\varphis{\varphi_{<}}
\def\varphil{\varphi_{>}}
\def\psibf{\boldsymbol{\psi}}
\def\Ome{\Omega}
\def\omeD{{\omega_D}} 
\def\bfOme{\boldsymbol{\Omega}} 
\def\Omebf{\boldsymbol{\Omega}} 
\def\lamb{\lambda}
\def\Lamb{\Lambda}
\def\sig{\sigma}
\def\Sig{\Sigma}
\def\sigp{{\sigma'}} 
\def\bfsig{\boldsymbol{\sigma}} 
\def\sigbf{\boldsymbol{\sigma}} 
\def\The{\Theta} 
\def\up{\uparrow}

\def\epsk{\epsilon_{\bf k}} 
\def\xik{\xi_{\bf k}} 
\def\txik{\tilde\xi_{\bf k}} 
\def\xip{\xi_{\bf p}} 
\def\xikq{\xi_{{\bf k}+{\bf q}}} 
\def\Ek{E_{\bf k}} 
\def\Ep{E_{\bf p}}
\def\Heff{\hat H_{\rm eff}}
\def\Hem{\hat H_{\rm em}}
\def\Hint{\hat H_{\rm int}}
\def\Hloc{\hat H_{\rm loc}}
\def\HMF{\hat H_{\rm MF}}
\def\Sem{S_{\rm em}}
\def\SMF{S_{\rm MF}} 
\def\SHF{S_{\rm HF}} 
\def\SRPA{S_{\rm RPA}} 
\def\Sint{S_{\rm int}} 
\def\Sloc{S_{\rm loc}}
\def\TN{T_{\rm N}} 
\def\TNHF{T^{\rm HF}_{\rm N}} 
\def\Zloc{Z_{\rm loc}} 
\def\ZMF{Z_{\rm MF}} 
\def\ZHF{Z_{\rm HF}} 
\def\ZRPA{Z_{\rm RPA}} 
\def\RPA{{\rm RPA}}
\def\loc{{\rm loc}} 
\def\pp{{\rm pp}}
\def\ph{{\rm ph}} 
\def\ch{{\rm ch}}
\def\sp{{\rm sp}} 
\def\qtf{q_{\rm TF}}
\def\epstf{\eps^{}_{\rm TF}} 
\def\epsrpa{\eps^{}_{\rm RPA}} 
\def\chinnzpp{\chi_{nn}^{0}{}\!\!\!''}

\def\half{\frac{1}{2}}
\def\dhalf{\dfrac{1}{2}}
\def\third{\frac{1}{3}} 
\def\quarter{\frac{1}{4}}

\def\qr{{\bf q}\cdot{\bf r}}
\def\wt{\omega t} 

\def\a{{\bf a}}
\def\b{{\bf b}}
\def\e{{\bf e}}
\def\f{{\bf f}}
\def\g{{\bf g}}
\def\h{{\bf h}}
\def\k{{\bf k}}
\def\l{{\bf l}}
\def\m{{\bf m}}
\def\n{{\bf n}} 
\def\p{{\bf p}} 
\def\q{{\bf q}}
\def\r{{\bf r}}
\def\t{{\bf t}}
\def\u{{\bf u}}
\def\v{{\bf v}}
\def\x{{\bf x}}
\def\y{{\bf y}} 
\def\z{{\bf z}} 
\def\A{{\bf A}}
\def\B{{\bf B}}
\def\D{{\bf D}} 
\def\E{{\bf E}} 
\def\F{{\bf F}} 
\def\H{{\bf H}}  
\def\J{{\bf J}}
\def\K{{\bf K}} 

\def\G{{\bf G}}
\def\L{{\bf L}}
\def\M{{\bf M}}  
\def\O{{\bf O}} 
\def\P{{\bf P}} 
\def\Q{{\bf Q}} 
\def\R{{\bf R}}
\def\S{{\bf S}}
\def\epsbf{\boldsymbol{\epsilon}}
\def\mubf{\boldsymbol{\mu}}
\def\nablabf{\boldsymbol{\nabla}}
\def\rhobf{\boldsymbol{\rho}}
\def\sigmabf{\boldsymbol{\sigma}} 
\def\Pibf{\boldsymbol{\Pi}}
\def\pibf{\boldsymbol{\pi}}

\def\para{\parallel}
\def\kpara{{k_\parallel}}
\def\kperp{{k_\perp}} 
\def\kperpp{{k_\perp'}} 
\def\qperp{{q_\perp}} 
\def\tperp{{t_\perp}} 

\def\w{\omega}
\def\wn{\omega_n}
\def\wnu{\omega_\nu}
\def\wp{\omega_p} 
\def\dmu{{\partial_\mu}}
\def\dl{{\partial_l}}  
\def\dt{\partial_t} 
\def\tdt{\tilde\partial_t}
\def\dk{\partial_k}
\def\tdk{\tilde\partial_k}
\def\dx{\partial_x}
\def\dy{\partial_y} 
\def\dtau{{\partial_\tau}}  
\def\det{{\rm det}} 
\def\Pf{{\rm Pf}}

\def\dsum{\displaystyle \sum}
\def\dint{\displaystyle \int} 
\def\intt{\int_{-\infty}^\infty dt} 
\def\inttp{\int_{-\infty}^\infty dt'} 
\def\intk{\int_{\bf k}} 
\def\intkd{\int \frac{d^dk}{(2\pi)^d}}
\def\intq{\int_{\bf q}} 
\def\intr{\int d^dr}  
\def\dintr{\displaystyle \int d^dr} 
\def\intrp{\int d^dr'}
\def\dinttau{\displaystyle \int_0^\beta d\tau}
\def\dinttaup{\displaystyle \int_0^\beta d\tau'}
\def\inttau{\int_0^\beta d\tau}
\def\inttaup{\int_0^\beta d\tau'}
\def\intx{\int d^{d+1}x} 
\def\inttaur{\int_0^\beta d\tau \int d^dr}
\def\intinf{\int_{-\infty}^\infty}
\def\dinttaur{\displaystyle \int_0^\beta d\tau \int d^dr}
\def\dintinf{\displaystyle \int_{-\infty}^\infty}
\def\intw{\int_{-\infty}^\infty \frac{d\w}{2\pi}}
\def\sumr{\sum_{\bf r}} 

\def\calA{{\cal A}} 
\def\calC{{\cal C}} 
\def\dt{\partial_t}
\def\calD{{\cal D}}
\def\calF{{\cal F}} 
\def\calG{{\cal G}}
\def\calH{{\cal H}}
\def\calI{{\cal I}}
\def\calJ{{\cal J}}
\def\calK{{\cal K}}
\def\calL{{\cal L}} 
\def\calN{{\cal N}}
\def\calO{{\cal O}}
\def\calP{{\cal P}}  
\def\calR{{\cal R}} 
\def\calS{{\cal S}}
\def\calT{{\cal T}}
\def\calU{{\cal U}}
\def\calX{{\cal X}} 
\def\calY{{\cal Y}} 
\def\calZ{{\cal Z}} 

\def\calFbf{{\bf F}}

\def\tT{{\tilde T}}
\def\talpha{{\tilde\alpha}}
\def\tdelta{{\tilde\delta}}
\def\teta{{\tilde\eta}} 
\def\tlamb{{\tilde\lambda}}
\def\tmu{{\tilde\mu}}
\def\tphibf{{\tilde\phibf}}
\def\trho{{\tilde\rho}}
\def\tvarphibf{{\tilde\varphibf}} 
\def\tw{{\tilde\omega}}
\def\twn{{\tilde\omega_n}}

\def\asinh{{\rm asinh}} 


\title{Thermodynamics of the two-dimensional XY model from functional renormalization } 

\author{P. Jakubczyk} 
\email{electronic address: pawel.jakubczyk@fuw.edu.pl}
\affiliation{Institute of Theoretical Physics, Faculty of Physics, University of Warsaw, Pasteura 5, 02-093 Warsaw, Poland}
\affiliation{Max Planck Institute for Solid State Research, Heisenbergstr. 1, 70569, Stuttgart, Germany}

\author{A. Eberlein} 
\email{electronic address: eberlein@physics.harvard.edu}
\affiliation{Department of Physics, Harvard University, Cambridge MA 02138, USA}

\date{April 21, 2016} 

\begin{abstract}
We solve the nonperturbative renormalization-group flow equations for the two-dimensional XY model at the truncation level of the (complete) second-order derivative expansion. We compute the thermodynamic properties 
in the high-temperature phase and compare the non-universal features specific to the XY model with results from Monte Carlo simulations. In particular, we study the position and magnitude of the 
specific heat peak as a function of temperature. The obtained results compare well with Monte Carlo simulations. We additionally gauge the accuracy of simplified nonperturbative renormalization-group treatments relying on 
$\phi^4$-type truncations. Our computation indicates that such an approximation is insufficient in the high-$T$ phase and a correct analysis of the specific heat profile requires  account of an infinite number of interaction vertices.

\end{abstract}
\maketitle

\section{Introduction}
It has since long been recognized \cite{Berezinskii70, Berezinskii71, Kosterlitz73, Kosterlitz74, Froehlich81} that the two-dimensional XY model undergoes a Kosterlitz-Thouless (KT) phase transition upon varying 
temperature $T$. This transition is peculiar in a number of 
respects: it is not accompanied by the appearance of long-range order (which is prohibited by the Mermin-Wagner theorem \cite{Mermin66}) and the free energy is a smooth ($C^{\infty}$ class) 
function of the thermodynamic parameters. Nonetheless, the low-$T$ phase displays long-range correlations and order-parameter stiffness. The latter exhibits a universal jump upon 
crossing the transition temperature $T_{KT}$. The correlation length is characterized by an essential singularity in the vicinity of the transition in the high-$T$ phase. A distinct nonuniversal feature of the 
$XY$ model is the 
pronounced, asymmetric peak of the specific heat as a function of $T$. The occurrence of this maximum is usually attributed to a rapid increase of entropy upon unbinding the vortex-antivortex pairs. 
The peak is located somewhat above the transition temperature $T_{KT}$. It is peculiar that on one hand the maximum is well separated from the asymptotic critical region, and, on the other, it occurs in a temperature 
range where the correlation length is still very large compared to the microscopic scale.

The Kosterlitz-Thouless transition is relevant in a number of physical contexts \cite{Chaikin_book} such as magnetism, liquid crystals, melting of two-dimensional ($d=2$) solids, 
superfluidity and superconductivity. 
Experimentally the KT-type behavior was observed in liquid-helium films \cite{Bishop78, Maps81, Maps82} and atomic gases \cite{Hadzibabic06, Tung10, Desbuquois12, Murthy15}.   

The universal aspects of the KT transition are conventionally described in the language of vortex-antivortex pair unbinding and a mapping to a Coulomb gas or sine-Gordon field theory. 
The predictivity of such formulations is typically restricted to the behavior in the vicinity of the transition, which makes it harder to access the non-universal, system specific properties, 
such as the critical temperature or the position, magnitude and width of the specific-heat peak.   

In the present work we develop and extend the description of the KT transition using the non-perturbative renormalization group (RG). 
We build upon earlier works \cite{Graeter95, Gersdorff01, Jakubczyk14}, which, however, were limited to $\phi^4$-type effective models and focused exclusively on universal aspects of the 
transition.
The formulation evades the explicit
introduction of vortices as degrees of freedom, and, in the form presented here, takes 
a microscopic spin system as the starting point. On the other hand, the approach captures the low-wavelength infrared (IR) asymptotics and respects the Mermin-Wagner theorem. 
In the present analysis we focus primarily on the relatively 
simple XY model on a square lattice, where 
the results obtained at different approximation levels of the RG framework can be compared to ample Monte Carlo (MC) data \cite{Tobochnik79, Himbergen81, Olsson95, Hasenbusch97, Hasenbusch05, Xu07, Komura12, Yu14}.
Observe however that the verification of the theoretical predictions of the KT theory by MC 
simulations has 
not always been conclusive even with respect to the most basic properties. For a critical discussion of these issues see Ref.~\cite{Hasenbusch05}. The present approach complements the MC in that it is formulated directly for infinite volume and does not 
invoke finite-size scaling theory. It also differs from the standard RG treatments in evading introduction of vortices or any expansions in powers of the order-parameter field. We show that the latter is crucial for a 
correct (even qualitatively) account of thermodynamics in the high-$T$ phase.

The nonperturbative RG is among the methods allowing for accurate computations of critical behavior in diverse systems. Its applicability is in addition by no means restricted to the vicinity of a phase transition. 
It has proven useful 
in a wide range of complex physical contexts. Examples include models with competing orders \cite{Metzner_review, Friederich11, Giering12, Eberlein14} or situations out of equilibrium \cite{Kloss12, Kloss14, Mesterhazy15}.
 The formalism by itself sheds light on fundamental aspects of critical phenomena (see e.g. \cite{Leonard15, Delamotte16}), leading to a genuine progress of the field. On the other hand, not so often do 
 the computations within this approach reach high-precision accuracy away from the critical region. The precise predictions also typically depend somewhat on the choice of 
regularization. Ref.~\cite{Machado10} shows that the critical temperature of the 3-dimensional Ising model may be calculated with the accuracy of around 1\%. Going beyond this precision level would require a substantial effort. 
The presently analyzed case of the 2-dimensional 
XY model is methodologically very distinct for at least two reasons: (1) The physics governing the vicinity of the phase transition is dominated by the anomalous dimension (which is negligible in $d=3$ for most purposes); (2) 
fluctuation effects are stronger (and lead to ultimate obliteration of long-range order) due to the presence of the Goldstone mode.  
 
Our framework automatically encodes the Mermin-Wagner theorem and is (upon slight modifications) extendable to more complex systems characterized by similar low-energy behavior at finite temperatures. These include quantum spins as well as interacting bosons or fermions 
in $d=2$. Such systems were already studied within simpler nonperturbative RG truncations, see e.g.~\cite{Krahl07, Floerchinger09, Rancon12, Rancon13, Rancon14, Strack14}. However, the latter are not sufficient to 
correctly account for nonuniversal features related to the KT 
transition - see Sec.~V. 
Before embarking on the more complex problems mentioned above, it is important to understand the merits and limitations of the method in situations where the results can be reliably compared to other approaches.  

\section{The XY model and the corresponding lattice field theory} 
The classical XY model on a lattice is defined by the Hamiltonian
\begin{equation}
\label{Hamiltonian}
\mathcal{H}\left(\{\vec{s_i}\}_{i=1}^{N}\right) = -\frac{1}{2}J_{ij}\vec{s_i}\vec{s_j}\;
\end{equation}
where $i,j\in\{1\dots N\}$ label the sites of the lattice, $\vec{s_i}\in \mathbb{R}^2$, $|\vec{s_i}|=1$, and the summation is implicit wherever the index appears exactly 
twice in a product expression. 

The corresponding partition function is given by 
\begin{equation} 
\label{Z_XY}
\mathcal{Z} = \sum_{\{\vec{s}\}}e^{-\beta\mathcal{H}}, \;\;\; \textrm{where} \;\;\; \sum_{\{\vec{s}\}} = \int \prod_{i}d\theta_i\;.
\end{equation}
Here $\beta^{-1}=k_BT$ and $\theta_i$ denotes the angle between the vector $\vec{s_i}$ and the $x$-axis, so that $\vec{s_i}\vec{s_j}=\cos(\theta_i-\theta_j)$ and 
$\theta_i\in [0,2\pi [$ for each $i$. 

In order to cast the problem of evaluating the partition function in the language of field theory, we employ the identity
\begin{equation} 
\label{identity}
e^{\frac{1}{2} A_{ij}\vec{s_i}\vec{s_j}}=\mathcal{N}^{-1}\int\prod_id\vec{\psi_i}e^{-\frac{1}{2}\left(\mathbb{A}^{-1}\right)_{ij}\vec{\psi_i}\vec{\psi_j}+\vec{s_i}\vec{\psi_i}}\;,
\end{equation} 
where the normalization factor is given by
\begin{equation}
\mathcal{N} = (2\pi)^N\det \mathbb{A}\;. 
\end{equation}
Here $\vec{\psi}_i$ is a two-dimensional vector attributed to the lattice site $i$. Eq.~(\ref{identity}) applies provided the matrix $\mathbb{A}$ is positive-definite. The non-positivity can 
 be cured by shifting the matrix by a constant diagonal term, which, in our setup, amounts to transforming the Hamiltonian Eq.~(\ref{Hamiltonian}) via 
\begin{equation}
\mathcal{H}\left(\{\vec{s_i}\}_{i=1}^{N}\right) \longrightarrow \mathcal{H}_c \left(\{\vec{s_i}\}_{i=1}^{N}\right) = -\frac{1}{2}J_{ij}\vec{s_i}\vec{s_j}-\frac{1}{2}c 
\vec{s_i}\vec{s_i}\;,
\end{equation}
i.e. shifting it by a constant equal $\frac{1}{2}Nc$. Specifying 
\begin{equation}
A_{ij} = \beta (J_{ij} +c\delta_{ij} ) 
\end{equation}
we apply Eq.~(\ref{identity}) to Eq.~(\ref{Z_XY}). The resulting expression for the partition function $\mathcal{Z}$ still involves the multiple integration over the spin variables $(\{\vec{s}\})$, which can now be explicitly performed    
\begin{equation}
\sum_{\{\vec{s}\}}e^{\vec{s_i}\vec{\psi_i}} = \left(2\pi\right)^N\prod_i I_0\left(|\vec{\psi_i}|\right)\;. 
\end{equation} 
Here $I_\alpha(x)$ denotes the hyperbolic Bessel function of first kind. This way we cast the partition function in the form
\begin{equation}
\mathcal{Z} = \left( \det \mathbb{A}\right)^{-1} \int \prod_i e^{-\frac{1}{2}\vec{\psi_i}(\mathbb{A}^{-1})_{ij}\vec{\psi_j}+\sum_i\ln I_0\left(|\vec{\psi_i}|\right) }\;. 
\end{equation}
In order to make all the temperature dependencies explicit, we rescale the interaction matrix $\mathbb{A}$ and the fluctuating field $\vec{\psi}$ according to: 
\begin{equation}
\label{change_of_var}
\tilde{A}_{ij} = \beta^{-1} A_{ij}\;,\;\;\;\;\;\;\; \vec{\phi_i} = \beta^{-\frac{1}{2}}\vec{\psi_i}\;,
\end{equation}
This way the partition function becomes expressed as 
\begin{equation}
\label{zet_rep}
\mathcal{Z} = \mathcal{D}\vec{\phi} e^{-\beta S[\vec{\phi}]},  
\end{equation}
with
\begin{equation}
\label{initial_action}
\beta S[\vec{\phi}] = \frac{1}{2} \vec{\phi_i}\left(\tilde{\mathbb{A}}^{-1}\right)_{ij}\vec{\phi_j}-\sum_i\log I_0(\beta^{\frac{1}{2}}|\vec{\phi_i}|) 
\end{equation}
and 
\begin{equation}
\label{measure}
\mathcal{D}\vec{\phi} = \left(\det \tilde{A}\right)^{-1} \prod_i d \vec{\phi_i}\;.
\end{equation}
Importantly, the change of variables of Eq.~(\ref{change_of_var}) removes temperature dependencies from the integration measure $\mathcal{D}\vec{\phi}$ as well as the kinetic term 
$\frac{1}{2}\vec{\phi_i}\left(\mathbb{A}^{-1}\right)_{ij}\vec{\phi_j}$ in the effective action $S[\vec{\phi}]$, and absorbs it fully into the local potential term 
$\log I_0(\beta^{\frac{1}{2}}|\vec{\phi_i}|)$. This aspect is crucial for the validity of the subsequent approximate RG procedure of Sec.~III and IV. The choice introduced in Eq.~(\ref{change_of_var}) 
differs from some standard conventions \cite{Amit_book}.

Eq.(\ref{zet_rep}-\ref{measure}) define the starting point for our computations. Specific lattice and interaction types may now be addressed by specifying the corresponding 
matrix $\mathbb{J}$.
Assuming translational invariance the kinetic term in $S[\vec{\phi}]$ is diagonalized with the Fourier transform:
\begin{equation}
 \frac{1}{2} \vec{\phi_i}\left(\tilde{\mathbb{A}}^{-1}\right)_{ij}\vec{\phi_j} = \frac{1}{2}\sum_{\vec{q}}\tilde{A}^{-1}_{\vec{q}}\vec{\phi}_{\vec{q}}\vec{\phi}_{-\vec{q}}\;,
\end{equation}
where 
\begin{equation}
 \tilde{A}_{\vec{q}} = c + J_{\vec{q}} = c + \frac{1}{N}J_{ij}e^{i\vec{q}(\vec{r}_i-\vec{r}_j)}\;.
\end{equation} 
This establishes the explicit form of the kinetic term. 
\subsection{Mean-field theory for ferromagnetic order} 
Assuming a form of $\mathbb{J}$ favoring ferromagnetic ordering, one identifies the mean-field free energy as the minimum of $S[\vec{\phi}]$. Restricting to uniform field 
configurations the mean-field equilibrium value of $|\phi|$ is given by 
\begin{equation}
\tilde{A}_0^{-1}|\vec{\phi}|-\frac{I_1(\beta^{1/2}|\vec{\phi}|)}{I_0(\beta^{1/2}|\vec{\phi}|)}\beta^{1/2} = 0\;.  
\end{equation}
The mean-field critical temperature is obtained by expanding the above around $|\vec{\phi}|=0$ up to terms linear in $|\vec{\phi}|$. This relates the critical temperature to 
$\tilde{A}_0$:
\begin{equation} 
\label{MF_temp}
k_B T_c = \frac {1}{2}\tilde{A}_0 \;. 
\end{equation} 
The corresponding critical exponents are classical and the critical temperature of Eq.~(\ref{MF_temp}) carries a strong, linear dependence on the parameter $c$ \cite{Amit_book}. Obviously, the resulting occurrence of long-range 
order at mean-field level contradicts the Mermin-Wagner theorem. In addition we observe that the mean-field free energy, and, in consequence, also the specific heat 
is zero in the high-temperature phase.
\subsection{Nearest-neighbor interactions}  
For the square lattice with nearest-neighbor interactions we obtain 
\begin{equation} 
\label{NNJ}
 J_{\vec{q}} = 2J\left[\cos(a q_x) + \cos(a q_y)\right]\;, 
\end{equation}
where $J$ is the nearest-neighbor coupling and $a$ denotes the lattice spacing. The latter will be put equal to 1 in all numerical calculations.   
\section{Nonperturbative RG}
The central idea of the nonperturbative renormalization group approach to equilibrium condensed-matter systems is to recast the problem of computing the partition function 
$\mathcal{Z}$ in a form of a (functional) differential equation. There exists a number of variants of this program. The presently applied formulation, developed by Wetterich \cite{Wetterich93}, 
relies on the concept of a flowing scale-dependent effective action $\Gamma_k[\vec{\phi}]$. This quantity continuously connects the microscopic action (in the present work given 
by Eq.~(\ref{initial_action})) with the full free energy $ F$ upon varying the flow parameter $k$. The latter is here taken to be an IR momentum cutoff scale. It serves 
to add a mass of order $\sim k^2$ to the fluctuation modes, effectively freezing their propagation for momenta $q<k$. Lowering the cutoff scale implies including modes of 
progressively lower momenta. For vanishing $k$ all fluctuation modes are included into the partition function and we find $\Gamma_k[\vec{\phi}]\longrightarrow \beta F[\vec{\phi}]$ 
as $k\to 0$. 
The variation of $\Gamma_k[\vec{\phi}]$ upon changing $k$ is governed by the flow equation \cite{Wetterich93}: 
\begin{equation}
\dk \Gamma_k[\vec{\phi}] = \half \Tr\llbrace \dk R_k\left(\Gamma^{(2)}_k[\vec{\phi}] + R_k\right)^{-1} \rrbrace ,
\label{rgeq}
\end{equation}
where $\Gamma^{(2)}_k[\vec{\phi}]$ denotes the second functional derivative of $\Gamma_k[\vec{\phi}]$. In Fourier space, the trace (Tr) sums over momenta and the field 
index $a\in \{1,2\}$. The quantity $R_k(q)$ is the momentum cutoff function added to the inverse propagator to freeze the fluctuations with momenta $q<k$. An exact solution 
of Eq.~(\ref{rgeq}) with the initial condition given by Eq.~(\ref{initial_action}) would imply finding the partition function $\mathcal{Z}$. This is not achievable, but the framework of 
Eq.~(\ref{rgeq}) offers a number of approximation schemes \cite{Berges02, Kopietz_book, Metzner_review, RG_book} going beyond those accessible within the more traditional approaches. 
\subsection{Derivative expansion}
In this work we apply the derivative expansion \cite{Berges02, RG_book, Canet03, Delamotte04} (DE) in which the symmetry-allowed terms in $\Gamma_k$ are classified according to the number of 
derivatives (or powers of $\vec{q}$ in momentum space). The most general expression at 
level $\partial^2$ (or $q^2$) reads: 
\begin{equation}
 \Gamma_k[\vec{\phi}] = \int d^2x \llbrace U_k(\rho) + \half Z_k(\rho) (\nablabf\vec{\phi})^2 + \quarter Y_k(\rho) (\nablabf\rho)^2 \rrbrace ,
 \label{DE}
\end{equation}
where $\rho = \half \vec{\phi}^2$. We impose restrictions neither on the effective potential $U_k(\rho)$ nor the gradient functions $Z_k(\rho)$, $Y_k(\rho)$, which are allowed to depend on the cutoff scale $k$. The 
occurrence of two gradient terms is due to the fact that the transverse and radial components of the field are characterized by different stiffness coefficients. 
We also observe here that the initial condition Eq.~(\ref{initial_action}) contains terms of all powers of $\rho$. In fact, the initial condition does not quite fit the ansatz of Eq.~(\ref{DE}) since the kinetic term involves functions 
of arbitrarily high  
order in $|\vec{q}|$. We come back to this point later on. Plugging the ansatz of Eq.~(\ref{DE}) into Eq.~(\ref{rgeq}) yields a projection of the Wetterich equation onto a set of three coupled 
non-linear partial differential equations describing the flow of  
$U_k(\rho)$, $Z_k(\rho)$ and $Y_k(\rho)$, which may be handled numerically. It is advantageous to perform a canonical rescaling of the flowing quantities by defining 
\begin{equation}
\begin{gathered}
\tilde U_k(\trho) = v_2^{-1} k^{-2} U_k(\rho), \quad \tilde Z_k(\trho) = Z_k^{-1} Z_k(\rho), \\ \tilde Y_k(\trho) = v_2 Z_k^{-2} Y_k(\rho) ,
\end{gathered}  
\label{dimless}
\end{equation} 
where $\trho=v_2^{-1} Z_k\rho$ and the factor $v_2^{-1}=8  \pi $ is conventional. The $k$-dependent constant $Z_k$ (wave-function renormalization) is defined by imposing the 
condition $\tilde Z_k(\trhor)=1$ where $\trhor$ is an arbitrary renormalization point on the rescaled grid. The scale-dependent anomalous dimension $\eta$ is then given by 
\begin{equation}
\eta_k = - k\dk \ln Z_k , 
\label{eta}
\end{equation}
and the physical anomalous dimension follows from $\eta=\lim_{k\to 0} \eta_k$. We refrain from quoting the lengthy explicit expressions for the flow equations. 
These are given in Ref.~\cite{Gersdorff01} and in the appendix of Ref.~\cite{Jakubczyk14}. 
The transition temperature $T_{KT}$ is extracted following Ref.~\cite{Jakubczyk14} by using the fact that the flowing minimum $\rho_{0,k}$ of the (nonrescaled) effective potential vanishes as 
$\rho_{0,k}\sim k^{\eta}$ in the low-$T$ phase. This is consistent with both the absence of the long-range order and algebraic decay of correlations governed by the anomalous dimension $\eta$. Since 
$Z_k\sim k^{-\eta}$ for $T<T_{KT}$, the minimum of the rescaled potential $\tilde{\rho}_{0,k} = v_2^{-1}Z_k \rho_{0,k}$ remains finite for $k\to 0$ as long as $T<T_{KT}$, and vanishes otherwise.

\subsection{Initial condition for the propagator}
The proposed approach relies on two approximations. First: the flowing effective action $\Gamma_k[\vec{\phi}]$ is parametrized by the ansatz of Eq.~(\ref{DE}). This implies retaining the most general $U(1)$-invariant  
form of the action but only up to terms of order $\partial^2$. In particular the local potential is allowed to contain arbitrarily high powers of $\rho$. Second: as we already remarked, the initial action of Eq.~(\ref{initial_action}) 
involves terms of higher order in $|\vec{q}|$ than $|\vec{q}|^2$. We, however, cast it in a form consistent with Eq.~(\ref{DE}) by expanding the dispersion in Eq.~(\ref{initial_action}) around $\vec{q}=0$. Physically this may be 
understood as ''smearing'' or coarse-graining the lattice structure ''by hand''. 
In a somewhat more 
subtle treatment one might split the flow into two stages. In the initial part ($k>a^{-1}$) of the flow hardly any renormalization of $Z_k(\rho)$ and $ Y_k(\rho)$ occurs, 
but the cosine dispersion may play a role. In the second stage ($k<a^{-1}$) the lattice no longer matters. 

An interesting alternative is provided by the lattice non-perturbative RG framework \cite{Machado10}, where the initial 
stage of 
the flow is overall bypassed, and the initial condition is not given by the bare action, but, instead, is computed from the local limit of decoupled sites. This program, however, places restrictions on the cutoff, which 
are most naturally fulfilled by a non-smooth Litim-type regulator \cite{Litim01}. This in turn renders the flow much less stable numerically. Such complications are most severe in $d=2$. We observe no signatures of numerical instabilities in our variant of 
approximation. In addition our calculation requires a significantly smaller field grid than that of Ref.~\cite{Machado10}. 

Our strategy to perform the $q$-expansion from the outset 
instead of the slightly more accurate treatments mentioned above also stems from the aim to develop a numerically modest and stable framework flexibly extendable to other contexts (such as interacting quantum gases in $d=2$). The present 
approximation allows to avoid any two-dimensional integrations in the flow equations. Also observe that the dominant contributions to all the integrals come from small momenta also for $k$ large, and, of course, upon 
reducing $k$ the approximation becomes progressively more accurate. By reasoning \textsl{a posteriori} our results suggest that the error from neglecting the cosine dispersion amounts to a shift of the critical 
temperature (see Sec. IV).

The second derivative matrix in the propagator at the beginning of the flow thus reads 
\begin{equation}
\frac{\delta^2 \beta S [\vec{\phi}]}{\delta \phi_{\vec{q_1}}^{\alpha_1}\delta \phi_{\vec{q_2}}^{\alpha_2} } = \delta_{\alpha_1, \alpha_2}\delta_{\vec{q_1}+\vec{q_2},0} \tilde{A}_{\vec{q_1}}^{-1} \;.
\end{equation}
We extract $Z_{k=k_0}(\rho)$ and $Y_{k=k_0}(\rho)$ from the $\vec{q_1}^2$ coefficients of the expansion of $\tilde{A}_{\vec{q_1}}^{-1}$ around $\vec{q_1}=0$. Here $k_0\gg a^{-1}$ is the initial cutoff scale. 
In fact, we obtain $Y_{k=k_0}(\rho)=0$ and $Z_{k=k_0}(\rho)=const>0$. We note that alternative to 
Eq.~(\ref{change_of_var}) rescalings of the fluctuating field and interaction matrix, such as that of Ref.~\cite{Amit_book}, generate both $Z_{k=k_0}(\rho)$ and $Y_{k=k_0}(\rho)$ dependencies in the initial condition.  

The momentum integrations in the flow equations are computed over a disc of radius $k_{UV} = \pi/a$. In a non-approximate treatment they should run over the Brillouin zone 
($]-\frac{\pi}{a},\frac{\pi}{a}]\times ]-\frac{\pi}{a},\frac{\pi}{a}] $ for a square lattice). The scale $k_{UV}$ is often identified with the scale $k_0$ where the flow is initiated. In fact these quantities are distinct, and, 
in principle $k_0$ should be taken infinite to assure that all fluctuations are frozen, so that the action Eq.~(\ref{initial_action}) is the correct starting point. In a practical numerical implementation 
we take $k_{UV}\ll k_{0} <\infty$ and assure that the inverse propagators at high scales are completely dominated by the cutoff term.

\subsection{Numerical solution}  
Numerical integration of the flow equations proceeds along the line of Ref.~\cite{Jakubczyk14}. There are however two important differences. Ref.~\cite{Jakubczyk14} used an effective 
$\phi^4$ action as a 
starting point, while here the initial condition follows from Eq.~(\ref{initial_action}) (see below). The other difference is that in the present calculation we extract thermodynamic quantities 
(specific heat in particular) related directly to the free energy, which is given by $U_{k\to 0}(\rho)$. While for the purposes of Ref.~\cite{Jakubczyk14} it was sufficient to compute the flow of the 
$\trho$-derivative of the rescaled potential $\tilde U_k'(\trho)$, here we additionally compute the flow of the (nonrescaled) potential $U_k(\trho)$. The corresponding flow equation, supplementing 
the flow equations given in Ref.~\cite{Jakubczyk14} reads: 
\begin{equation} 
\begin{gathered}
k^{-1}\partial_k U(\tilde{\rho}) = \tilde{\rho}\eta \tilde{U}_k'(\tilde{\rho})- \\ \int dx \left(\eta x r(x) +2x^2r'(x)\right) \left(\tilde{G}_L(x,\tilde{\rho}) + \tilde{G}_T(x,\tilde{\rho})\right) \;,
\end{gathered}
\end{equation} 
where $x=q^2/k^2$ and the dimensionless $\tilde{\rho}$-dependent longitudinal and transverse inverse propagators are given by 
\begin{equation}
\begin{split}
\tilde G_{\rm L}^{-1}(x,\tilde{\rho}) &= x[\tilde Z_k(\tilde{\rho}) + \trho \tilde Y_k(\tilde{\rho}) + r(x)] + \tilde U_k'(\tilde{\rho}) + 2\trho \tilde U_k''(\tilde{\rho})  \\    
\tilde G_{\rm T}^{-1}(x,\tilde{\rho}) &=  x[\tilde Z_k(\tilde{\rho}) + r(x)] + \tilde U_k'(\tilde{\rho})  \;.
\end{split}
\end{equation}

A reliable calculation demands high numerical accuracy. This is because we solve the flow equations for a set of initial conditions parametrized by temperature $T$ and subsequently 
evaluate the entropy and the specific heat by numerically computing the first two derivatives of the result for the free energy with respect to $T$. 

We also observe that, since we explore the high-temperature phase and the vicinity of the phase transition in the low-$T$ phase, the principal problem of encountering the pole of the flow equations 
described in Ref.~\cite{Jakubczyk14} is irrelevant here. 

In the practical numerical solution we employ the smooth Wetterich cutoff 
\begin{equation} 
R_k(\vec{q}) = Z_k \vec{q}^2 r (\vec{q}^2/k^2), \qquad r(x) = \frac{\alpha}{e^x-1} \;.
\label{Rdef}
\end{equation}
The inclusion of the wave-function renormalization in $R_k(\vec{q})$ is a requirement for the possibility of obtaining scale-invariant solutions. The parameter $\alpha$ is in principle arbitrary.  
Ref.~\cite{Jakubczyk14} rose the question of the existence of exact (functional) fixed points of the flow depending on its value. The present analysis is performed at fixed $\alpha=2.0$ which is close to the 
''optimal'' value 
in the immediate vicinity of the transition. We refer to \cite{Jakubczyk14} for an extensive discussion of this issue. 
\section{Numerical Results}
For the nearest-neighbor XY model $J_{\vec{q}}$ is given by Eq.~(\ref{NNJ})
and it follows that the initial condition for $Z_k(\rho)$ and $Y_k(\rho)$ reads
\begin{equation}
Z_{k_0}(\rho) = Z_0 = \frac{J a^2}{(c+4J)^2}\;,\;\;\;\;\; Y_{k_0}(\rho)=0\;, 
\end{equation}
while the initial effective potential is given by 
\begin{equation}
U_{k_0}(\rho) = U_0(\rho) =  \frac{1}{(c + 4J)}\rho - \log I_0 \left(\sqrt{2\rho\beta}\right) \;.
\end{equation}
We observe that both $U_0(\rho)$ and $Z_0$ depend on the arbitrary parameter $c$. In fact, as we already pointed out, the mean-field transition temperature of Eq.~(\ref{MF_temp}) carries a strong $c$-dependence. Adding 
fluctuations by the 
non-perturbative RG flow drastically reduces this dependence, but does not remove it completely, as discussed below. Also observe that 
(at least formally) the above expressions for $U_0(\rho)$ and $Z_0(\rho)$ make sense for arbitrary non-negative values of $c$. On the other hand, for the present case of nearest-neighbor interactions, the  
matrix $\mathbb{A}$ is positive-definite for $c>4$. 
\subsection{Critical temperature}
\begin{figure}
\centerline{\includegraphics[width=8.5cm]{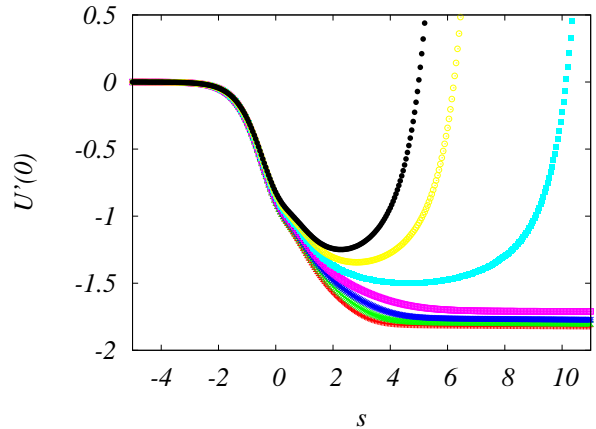}}
\caption{(Color online) Exemplary flows of the (rescaled) derivative of the effective potential at $\tilde{\rho}=0$ for a range of temperatures containing $T_{KT}$. For $T>T_{KT}$ the minimum of the rescaled potential 
($\tilde{\rho}_0$) hits zero at a finite scale $k$, and $\tilde{U}'(0)>0$. On the other hand, for $T\leq T_{KT}$ the flowing couplings (including $\tilde{U}'(0)$) attain fixed-point-like behavior. The lowest curve 
corresponds to $T\approx 0.95 T_{KT}$, the highest one to $T\approx 1.05 T_{KT}$. } 
\label{flow_plot} 
\end{figure}

The critical temperature $T_{KT}$ is estimated by following the flow of the minimum of the (rescaled) effective potential. This quantity reaches zero at a finite scale $k>0$ for the system 
in the high-$T$ phase, and attains an (approximate) fixed-point in the KT-phase. Equivalently, one may inspect the evolution of the anomalous exponent $\eta_k$, vanishing in the high-$T$ phase for $k$ sufficiently 
small and attaining a ''plateau'' otherwise. The procedure follows Ref.~\cite{Jakubczyk14} and is illustrated in Fig.~\ref{flow_plot} where we plot $\tilde{U}'(0)$ as a function of $s=-\log\left(k/k_0\right)$. 
Also note that the method of estimating $T_{KT}$ is different from the MC, which typically employs a 
fit of the theoretical formulae for the correlation length and susceptibility to the simulation data. 
In Ref.~\cite{Machado10} only a very weak dependence of the critical temperature on the parameter $c$ was found in the case of the Ising model in three dimensions. 
In the present case we observe a monotonous dependence  
of the KT temperature on the parameter $c$ ranging between $0.91 J/k_B$ for $c=4$ and $1.02 J/k_B$ for $c=8$. The dependence of $T_{KT}$ on $c$ slowly ceases at larger values of $c$. Large values of $c$ are however 
very unpractical because 
$Z_0\sim c^{-2}$ becomes very small. Our estimate of $T_{KT}$ may be compared to the MC results which give 
$T_{KT}\approx 0.89J/k_B$. The lattice version of nonperturbative RG yielded the estimate $0.9<T_{KT}/J<1$ \cite{Machado10}. 

We believe that the mechanism responsible for the annihilation (or significant reduction) of the $c$-dependence of the critical temperature is related to the fact that even though the position of the 
minimum of $U_0(\rho)$ carries a strong $c$-dependence, the minimum $\tilde{\rho}_0$ of the initial rescaled potential $\tilde{U}_0(\tilde{\rho})$ shows only a very weak sensitivity to the variation of $c$. On the other hand, the 
profile of $\tilde{U}_0(\tilde{\rho})$ for $\tilde{\rho}<\tilde{\rho}_0$ does depend on $c$. The dependence of $T_{KT}$ on $c$ should be efficiently eliminated by the flow in situations, where the essential features of the 
flow are captured by the behavior of the effective action around $\tilde{\rho}=\tilde{\rho}_0$. The mechanism is expected to be less efficient otherwise. This condition is better fulfilled in $d=3$.
Our procedure of performing the $q$-expansion from the beginning is also of relevance for the results for $T_{KT}$. The dependence of our estimate of $T_{KT}$ on $c$ is an unpleasant feature and we perceive it as 
a deficiency of the present approach. It is possible to choose $c$ so that we obtain $T_{KT}$ in precise agreement with MC, but this is not what we aim at. 
We note however that the $c$-dependence of $T_{KT}$ is by far weaker than at mean-field level. In addition, the thermodynamic quantities discussed below are insensitive to the choice of $c$ provided they are 
computed relative to $T_{KT}$. This suggests that the error related to our approximation is absorbed by a shift of $T_{KT}$, leaving other thermodynamic quantities hardly affected.   

\subsection{Entropy and specific heat} 
\begin{figure}
\centerline{\includegraphics[width=8.5cm]{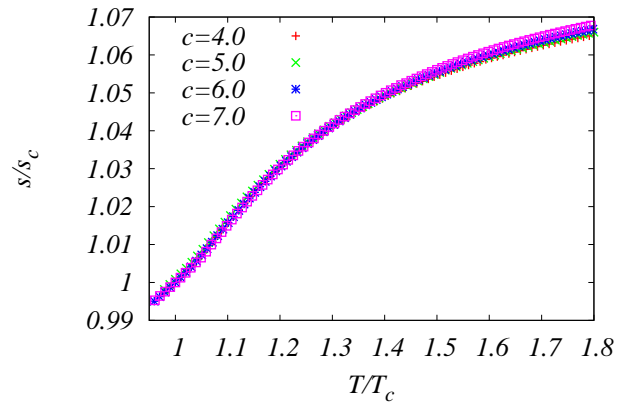}}
\caption{(Color online) Entropy as a function of the reduced temperature for a sequence of values of $c$. The curves collapse once scaled by the critical temperature $T_c=T_{KT}$ and the corresponding entropy density $s_c$. } 
\label{entropy_plot} 
\end{figure}

\begin{figure}
\centerline{\includegraphics[width=8.5cm]{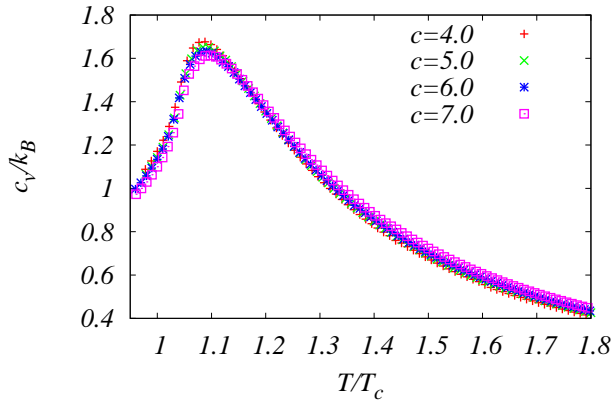}}
\caption{(Color online) Specific heat as function of the reduced temperature for a sequence of values of $c$. The location and magnitude of the peak agree well with MC and tensor RG (see the main text). } 
\label{c_v_plot}  
\end{figure}

We proceed by evaluating the entropy at zero magnetic moment (or, equivalently, zero magnetic field), which, by elementary thermodynamics follows from $S(T,N) = Ns(T) = - \frac{\partial F}{\partial T}$. The free energy 
$F(T,\phi=0,N)$ is obtained from the integrated flow via $F=k_B T\lim_{k\to 0} U_k(0)$. It is also possible to extend the analysis to non-zero fields since the magnetic field, the order parameter field and 
free energy are related by 
$h = k_B T\lim_{k\to 0} \partial_{|\phi|} U_k(|\phi|)$. By computing the flow for different $T$ we extract the free energy profiles $U(\rho)$ for a range of temperatures and subsequently evaluate the (discrete) 
derivative with respect to $T$. The results are shown in Fig.~(\ref{entropy_plot}). We observe a collapse of the curves computed for different $c$ if the variables are scaled by the critical values. The entropy is a 
positive, 
monotonously increasing function of temperature, as expected from the principles of thermodynamics. 
The signatures of the transition are not visible in the $T$-derivatives of the thermodynamic potential (as is expected from the $KT$-theory and also consistent with the 
results of simulations). 
In the next step we evaluate the specific heat (at zero magnetization). This is given by 
\begin{equation}
c_v = T\frac{\partial s}{\partial T}=-\frac{T}{N}\frac{\partial^2 F}{\partial T^2}\;. 
\end{equation}
The results obtained for different choices of $c$ are plotted in Fig.~(\ref{c_v_plot}). Again, the dependence on $c$ does not occur once we use the reduced variable. The pronounced maximum shows an asymmetry similar to those 
found in MC. The peak is located around $T_p\approx 1.1T_{KT}$ and reaches up to $c_v^m\approx 1.6k_B$. Both these quantities are close to the MC  and tensor RG results. 
More specifically (see e.g.~\cite{Tobochnik79, Xu07, Yu14}), the MC peak is located at 
$T_p^{MC}\approx 1.15 T_c$ and reaches up to $c_v^{m(MC)}\approx 1.55k_B$. Note however, that the free energy plotted in Ref.~\cite{Yu14} has positive slope, and the entropy obtained therein is 
negative. The reasons for this are not clear to us. The level of agreement of $T_{KT}$ and $c_v(T)$ between the MC and tensor RG results reported in \cite{Yu14} is very high.  

It is striking that the rich thermodynamic structure described in this section emerged via the functional RG flow from the mean-field free energy, which is trivially 
equal zero in the high-$T$ phase (see Sec.~II A).

\subsection{Equation of state}
\begin{figure}
\centerline{\includegraphics[width=8.5cm]{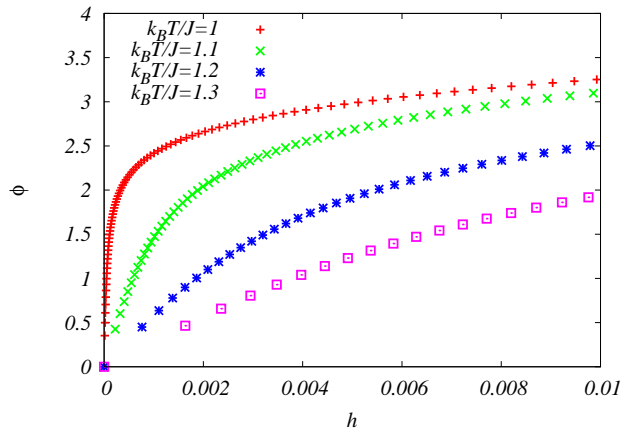}}
\caption{(Color online) Dependence of the order-parameter on magnetic field for a sequence of temperatures. The presented calculation was performed at $c=4$. } 
\label{mag_field_plot} 
\end{figure}

The magnetic field $\vec{h}$ at given $\vec{\phi}$ is extracted from the definition
\begin{equation}
\vec{h}=\frac{\partial F}{\partial \vec{\phi}} = k_B T \lim_{k\to 0}\frac{\partial U_k(\rho)}{\partial \vec{\phi}}\;, 
\label{hphi}
\end{equation}
which yields the equation of state $\vec{h}(T,\vec{\phi})$. The isothermal susceptibility at zero field is given as the derivative 
\begin{equation}
\chi^{-1} (T) = \frac{\partial h}{\partial \phi}|_{\phi=0} = k_BT v_2^{-1}\lim_{k\to 0}\left(Z_k \frac{\partial U_k(\tilde{\rho})}{\partial \tilde{\rho}}\right)|_{\tilde{\rho}=0}\;.  
\end{equation}
and becomes very large upon lowering temperature towards $T_{KT}$. The dependence $\phi(h)$ is shown in Fig.~\ref{mag_field_plot} for a sequence of temperatures.  
\section{Remarks on the $\phi^4$ truncation} 
It is useful to compare the above calculation to a much simpler treatment relying on the $\phi^4$-type expansion. It is natural to invoke a largely simplified ansatz for the effective potential 
\begin{equation}
U_k(\rho) = \frac{\lambda_k}{2}\left(\rho-\rho_{0,k}\right)^2 + \gamma_k\;, 
\label{phi4_1}
\end{equation}
and also restrict $Z_k(\rho)$ and $Y_k(\rho)$ to flowing couplings corresponding to the functions evaluated at the potential minimum ($\rho_0$). The problem may then be cast onto a set of five coupled ordinary 
differential flow equations for the couplings $\rho_{0,k}$, $\lambda_k$, $\gamma_k$, $Z_k$ and $Y_k$. The initial condition for the potential is extracted by expanding the effective potential in 
Eq.~(\ref{initial_action}) around its minimum. The ansatz Eq.~(\ref{phi4_1}) makes sense for $Z_k\rho_{0,k}>0$. Once the flow crosses over into the regime with $Z_k\rho_{0,k}=0$ one switches to the 
parametrization suitable for the high-temperature phase 
\begin{equation}
 U_k(\rho) = \frac{\lambda_k}{2}\rho^2 + \delta_k\rho + \gamma_k\;.
\label{phi4_2}
\end{equation}
The free energy may then be extracted from $\lim_{k\to 0}\gamma_k$. In fact, a very similar truncation (neglecting $Y_k$ and $\gamma_k$) was employed in Ref.~\cite{Graeter95} and yielded a plausible picture
of the KT transition. 

We have solved the above mentioned set of flow equations and compared the results to those obtained within the complete derivative expansion in Sec.~IV. Even though the estimate of the critical 
temperature $T_{KT}$ is in a reasonable range, the $\phi^4$ approximation badly fails for the thermodynamic quantities. In fact the obtained free energy $F(T)$ is not a concave function of temperature, yielding, 
for example, a negative specific heat in a range of temperatures. The reason for this becomes clear after inspecting Eq.~(\ref{initial_action}). Expanding the effective potential in $\rho$ implies uncontrolled 
dropping of temperature dependencies, which, as turns out, leads to a drastic deformation of the result.  
\section{Summary and outlook}
We have solved the non-perturbative RG flow equations for the two-dimensional XY model at the approximation level of the complete second order derivative expansion. 
From the obtained free energy $F(T,\phi,N)$ we computed the non-universal thermodynamic 
properties in the the high-temperature phase.  
Wherever possible, we compared the results to Monte Carlo simulations. We found satisfactory agreement for the entropy and specific heat. In particular, the location and magnitude of the specific heat peak relative 
to $T_{KT}$ compare very well to MC data. 
This is one of the few RG-based computations for this quantity in 
this model. As we pointed out, the specific heat peak occurs in a regime which on one hand is off the asymptotic critical region, and, on the other, is characterized by large correlation length. Such a situation is 
somewhat atypical. An interesting RG calculation was performed in Ref.~\cite{Yu14} within the tensor RG framework, which may be viewed as a reincarnation of the ideas of direct real-space 
coarse-graining scheme. However, it is not clear to what extent that approach can be generalized to other systems. 

Our estimate of the 
Kosterlitz-Thouless temperature is not far from the correct value, however the present method cannot serve as a high-precision tool in this case. We argue that the location of the transition is the quantity that is 
most strongly affected by the approximations, in particular by the relatively simple treatment of the dispersion at the initial stages of the flow. 
As we pointed out, the thermodynamic functions become insensitive to the arbitrary parameter $c$ of the Hamiltonian upon scaling by $T_{KT}$. 

We compared the full derivative expansion to a simplified treatment 
invoking vertex expansion ($\phi^4$-type theory), which is commonly applied in different contexts. The latter framework turns out not to be sufficient for computing the non-universal thermodynamic quantities, 
since it truncates relevant temperature dependencies in the neglected vertices. 

The present calculation bridges a microscopic model with macroscopic thermodynamics via the functional flow equation, accounting for the low-energy asymptotics specific to two-dimensional systems with $U(1)$ symmetry. 
It will now be natural and interesting to perform analogous studies of systems characterized by similar infrared physics, including interacting quantum gases in $d=2$.

\begin{acknowledgments}
We are grateful to Nicolas Dupuis and Walter Metzner for useful discussions. We also thank Walter Metzner for reading the manuscript and a number of valuable remarks. 
PJ acknowledges funding from the Polish National Science Center via grant 2014/15/B/ST3/02212. AE acknowledges support from the German National Academy of Sciences
Leopoldina through grant LPDS~2014-13.  
\end{acknowledgments}

\end{document}